\documentclass[aip,apl]{revtex4-1}
\usepackage{amsfonts}
\usepackage{amsmath}
\usepackage{amssymb}
\usepackage{graphicx}%
\usepackage{multirow}
\usepackage{bm}%
\usepackage{color}%
\begin{document}

\title{Inverse effect of magnetostriction in magnetoelectric laminates}

\author{M. Auslender}%
\author{E. Liverts}%
\author{B. Zadov}%
\author{A. Elmalem}%
\author{A. Zhdanov}%
\author{A. Grosz}%
\author{E. Paperno}%
\affiliation{Ben-Gurion University of the Negev, P.O. Box 653, Beer-Sheva 84105, Israel}%
\date{\today}

\begin{abstract}
We introduce the notion of inverse effect of magnetostriction for magnetostrictive-piezoelectric heterostructures and study this effect theoretically and experimentally. It is shown that the inverse effect of magnetostriction may crucially contribute to the mechanism of magnetoelectric coupling. It is shown that the studied effect essentially modifies the saturation magnetostriction of the whole structure as compared to its magnetic phase bulk and also induces an additional magnetic anisotropy. Our consideration provides useful insight into the fundamental issue of strain-mediated magnetoelectric coupling. Understanding this effect may lead to its utilization in original experimental concepts and the improvement of the ME coupling.
\end{abstract}

\pacs{Valid PACS appear here}
\maketitle

Magnetoelectric (ME) laminates that comprise bonded parallel magnetostrictive and piezoelectric layers are important subfamily of two-phase multiferroics \cite{Scott}. In these heterostructures, the indirect strain-mediated ME coupling results from an interplay of magnetostriction in the magnetic layer, piezoelectricity in the piezoelectric layer, and the bonding. So far, {\em inverse effect of magnetostriction} was meant as a change in the magnetic state of magnetostrictive media subjected to external stresses, being strong when the magnetoelastic and magnetic anisotropy energies are comparable \cite{bozorth,Chikazumi,magnet,Miyauchi,livingston,Zheng}. Recently, similar effect of residual stresses in a magnetostrictive film on a compliant substrate was observed \cite{Li}.  In the ME laminates subjected to an external magnetic field, internal stresses emerge as a result of mechanical compliance of the phases. As regards the magnetic phase, these stresses will drive the inverse effect of magnetostriction in the usual sense. In addition due to the hidden stresses, the whole laminate will deform upon increasing the magnetic field, quite differently than free standalone magnetic layer. We define these two phenomena altogether, as the inverse effect of magnetostriction for ME laminates. Though the strain-mediated ME coupling in the laminates has been intensively studied, \cite{Scott,srin,nan} and references therein, no specific attention has been paid to the laminate derived inverse magnetostriction effect. Even more, we believe that just missing the inverse effect of magnetostriction from the scope is a physical reason behind drastic theoretical overestimation of the ME coupling as compared to its experimental value \cite{{laletin},{liverts1}}. The present work reports the study of the inverse effect of magnetostriction for the ME laminates. Restricting ourselves to a range of magnetic field near the magnetic saturation allows us to combine physical rigor with relatively simple treatment. Our consideration concerns the polycrystalline magnetic phase, which is consistently taken into account, though the effect under study works also for crystalline magnetic one. We prove that when the phase has low magnetic anisotropy, the inverse effect of magnetostriction is crucial and may change the ME coupling by an order of magnitude as compared to the estimate obtained from piezomagnetic coefficient for the magnetic phase alone. The proposed account of the inverse effect of magnetostriction also provides emergent experimental concepts in the field of ME heterostructures.

Magnetostrictive polycrystalline layers, whose structural components are sufficiently small single-crystal grains, may be regarded as an isotropic body in terms of both elastic and magnetostrictive properties, especially in the range close to the magnetization saturation (since we are dealing with the strains and magnetization distribution over the spatial regions that are large compared to the grains size)~\cite{landau_el,Chikazumi,magnet}. In the cases where the anisotropy of the elastic properties is relatively small, and magnetostriction constants are of the same sign (e.g. Ni), the magnetostriction constants and elastic modules of the isotropic body can be accurately found from those of the single crystal by straightforward calculating the strains in each grain and then averaging them over the grains ~\cite{landau_el,magnet}.

For a strained cubic crystal, the magnetoelastic energy expressed for each domain with a local frame $x,\;y,\;z$ in terms of the strain  $u_{ik}$,  magnetization $\mathbf{M}$,  $\mathbf{m}=\mathbf{M}/M_s$ ($M_s$ is the saturation magnetization), and two independent coefficients $a_1$, $a_2$, is given by
\begin{equation}
\mathcal{U}_{\rm m-el}=a_{1}\left(u_{xx}m_{x}^{2}+
u_{yy}m_{y}^{2}+u_{zz}m_{z}^{2}\right)
+a_{2}\left(u_{yz}m_{y}m_{z}+
u_{xz}m_{x}m_{z}\\
+u_{xy}m_{x}m_{y}\right).
\label{F_me}
\end{equation}
The change in the deformation resulting from a change in the magnetization direction and from persistent stress $\sigma_{ik}$ are obtained by minimizing energy $\mathcal{U}^m_{\rm el}+\mathcal{U}_{\rm m-el}$, where $\mathcal{U}^m_{\rm el}$ is the elastic energy involving three independent elastic constants ($s^m_{11}$, $s^m_{12}$, $s^m_{44}$), is equal to
\begin{equation}
u_{xx}=s^m_{11}\sigma_{xx}+s^m_{12}(\sigma_{yy}+\sigma_{zz})+\frac{3}{2}\lambda_{100}\left(m_x^2-\frac{1}{3}\right),
\;u_{xy}=\frac{s^m_{44}}{2}\sigma_{xy}+3\lambda_{111} m_xm_y,
\label{strain}
\end{equation}
where $s^m_{44}a_{2}=-3\lambda_{111}$, $2(s^m_{12}-s^m_{11})a_{1}=3\lambda_{100}$, and similarly for the other components. According to \cite{landau} the above strain tensor presents all the magnetostriction effects in the cubic crystals.

Let $n_x$, $n_y$, $n_z$ be the cosines of a fixed direction $\mathbf{n}$ in the local cubic crystallographic frame of a crystallite, which serves as the axis $1$ of a laboratory frame with the axes $1$, $2$, $3$. 
Then, we adopt the following transformation between the local and laboratory frame
\begin{equation}
x_{1} = xn_x+yn_y+ zn_z, \;\;
x_{2} = x\frac{-n_y\cos\psi+n_xn_z\sin\psi}{\sqrt{1-n_z^2}}+y\frac{n_x\cos\psi+n_yn_z\sin\psi}{\sqrt{1-n_z^2}}-z\sqrt{1-n_z^2}\sin\psi.
\label{trans&m^2bot}
\end{equation}
The expression for $x_3$ can be obtained from the equation for $x_2$ under the replacement $\psi\rightarrow \psi-\pi/2$.
Assume that within the magnetostrictive layer the strains, stresses, and magnetization are nearly uniform and close to their saturation values. Then using Eq.~(\ref{trans&m^2bot}), averaging the strains in Eq.~(\ref{strain}) over all the directions of $\mathbf{n}$ (using the following computations $\langle n_i^4\rangle=1/7$, $\langle n_i^8\rangle=1/9$ etc.), $0\leq \psi\leq 2\pi$, and comparing the results with the appropriate expressions for an isotropic medium, we find that the elastic properties can be described by the Young's modulus $E$ and Poisson's ratio $\nu$, given by
\begin{equation}
\frac{1}{E}=\frac{3}{5}s^m_{11}+\frac{1}{5}\left(2s^m_{12}+s^m_{44}\right),\quad 2\nu=-\frac{2s^m_{11}+8s^m_{12}-s^m_{44}}{3s^m_{11}+2s^m_{12}+s^m_{44}},
\end{equation}
and the magnetostriction  can be described by a single coefficient $\lambda_s $, as in truly isotropic material.

To describe the inverse effect of magnetostriction in a bonded thin magnetostrictive-piezoelectric layered structure, we will define the in-plane coordinate axes as $1,\;2$ and the off-plane axis as $3$. Under an external magnetic field along the $1$ direction ($H_1=H,\;H_2=H_3=0$), the magnetostrictive layer shrinks or expands along the $1$, $2$ directions. If there is no slippage or fracture in the bonded trilayer, the magnetostrictive strain applies an in-plane constraint on the compliant piezoelectric layer which also can either shrink or stretch to minimize its energy $\mathcal{U}^p_{\rm el}+\mathcal{U}_{\rm e-el}$ and can help the piezoelectric to have induced polarization along the 3 axis according to its phase diagram. Here $\mathcal{U}_{\rm e-el}$ is the usual electro-elastic energy of the piezoelectric, which we take in a strain-charge form \cite{landau,{mason}}. For the 1-2 plane considered to be infinite, the uniform stresses corresponding to the uniform strain, are then imposed on each of the two phases to maintain displacement compatibility and equilibrium. These stresses are expressed via the layer's thicknesses and piezostrictive, magnetostrictive, and elastic properties; the off-plane stress components are neglected due to the small thickness and the fact that the trilayer structure prevents bending.
Taking into account that the forces per unit area exerted by the stresses in the magnetostrictive layers are balanced by those caused by the stresses in the piezoelectric and considering open-circuit condition, we obtain at the magnetic saturation
\begin{equation}
\sigma_{11}=-\frac{t_p\lambda_s}{2(s_{11}-s_{12})}\frac{2s_{11}+s_{12}+3t_md_{31}^2/\varkappa}{s_{11}+s_{12}-2t_md_{31}^2/\varkappa},\;\;\;
\sigma_{22}=\frac{t_p\lambda_s}{2(s_{11}-s_{12})}\frac{s_{11}+2s_{12}+3t_md_{31}^2/\varkappa}{s_{11}+s_{12}-2t_md_{31}^2/\varkappa},
\label{sigma_1&2&6}
\end{equation}
where $t_p$ is the piezoelectric thickness, $t_m=1-t_p$, $s_{11}= t_pE^{-1}+t_ms^p_{11}$, $s_{12}=-\nu t_pE^{-1}+t_ms^p_{12}$, $s^p_{ik}$, $d_{31}$ and $\varkappa$ are the piezoelectric elastic constants, piezoelectric coefficient, and out-of-plane dielectric permittivity, respectively. As it recently has been argued \cite{liverts} and also confirmed numerically \cite{zadov} by a finite element method, the expressions in Eq.(\ref{sigma_1&2&6}) are leading asymptotic in terms of the lateral-to-transverse dimensions ratios tending to infinity. The simulations \cite{zadov} for finite-lateral sizes' structures have shown deviation from Eq.(\ref{sigma_1&2&6}) only in transition layers near the edges, where the physical fields are strongly inhomogeneous.

One of the outcomes, most important for applications, is that a significant change in the averaged deformation occurs in the laminate as compared to the deformation in the bare magnetostrictive phase. The average strains at $H\gg M_s$, resulting from the magnetostriction, are
$\langle u_{11}\rangle =(\sigma_{11}-\nu\sigma_{22})E^{-1}+\lambda_s$ and $\langle u_{22}\rangle=(\sigma_{22}-\nu\sigma_{11})E^{-1}-\lambda_s/2$.
Then the calculated magnetostriction $\lambda_\theta$ at an angle $\theta$ to the magnetization is given by
\begin{equation}
\lambda_\theta=\frac{\delta l_\theta}{l}=\frac{1-\nu}{2E}(\sigma_{11}+\sigma_{22})+\frac{\lambda_s}{4}+
\left[\frac{1+\nu}{2E}(\sigma_{11}-\sigma_{22})+\frac{3\lambda_s}{4}\right]\cos2\theta.
\label{str}
\end{equation}
The measurement of $\lambda_\theta$ vs $\theta$ allows one to extract the lamination stresses and compare them with those theoretically predicted by Eq.(\ref{sigma_1&2&6}).

At elastic equilibrium between the bonded magnetostrictive and piezoelectric layers, the total energy of the former $\mathcal{U}^m\left(\mathbf{m},\mathbf{H}\right)$ is eventually obtained by substituting the equilibrium strains, see Eq.(2), into the expression of $\mathcal{U}^m_{\rm el}+\mathcal{U}_{\rm m-el}$. This yields
$\mathcal{U}^m = \mathcal{U}^m_{0}+\Delta \mathcal{U}^m$, where
 $\mathcal{U}^m_{0}$ is sum of the Zeeman energy and the energy of crystallographic magnetic anisotropy of the bare magnetostrictive phase at {\em zero strain}, and $\Delta\mathcal{U}^m$ is given by
\begin{equation}
\Delta\mathcal{U}^m\left(\mathbf{m}\right) = \Delta\mathcal{U}^m_{c}\left(\mathbf{m}\right)-\frac{3}{2}\lambda_{100}(m_x^2\sigma_{xx}+m_y^2\sigma_{yy}+m_z^2\sigma_{zz})
-3\lambda_{111}(m_xm_y\sigma_{xy}+m_xm_z\sigma_{xz}+m_ym_z\sigma_{yz}).
\label{deltaU^m}
\end{equation}
Here $\Delta\mathcal{U}^m_{c}$ is the cubic anisotropy energy induced by the magnetostriction in a bulk ferromagnetic at zero stress (coinciding with expressions given in the literature, see e.g. \cite{Chikazumi}), while the rest terms represent the magnetic anisotropy due to the stress induced by the lamination, which is of our primary interest. When the stress is given and is independent of $\mathbf{m}$, the stress-induced anisotropy is of uniaxial type. Although $|\Delta\mathcal{U}^m\left(\mathbf{m}\right)|<|\mathcal{U}^m_{0}\left(\mathbf{m},0\right)|$, the induced anisotropy is {\em not negligible}.
In what follows, we consider the magnetization curve of the laminate near the magnetic saturation. To this end, we search the minimum of the above magnetic energy (normalized to $\mu_0M_s^2$), keeping strict condition $|\mathbf{m}|=1$ with the use a perturbation theory in small parameter $\varepsilon =M_s/H$. Let us write the minimized function in the form
\begin{equation}
\frac{\beta}{4}\left(m_x^4+m_y^4+m_z^4\right)+\frac{\Delta\mathcal{U}^m\left(\mathbf{m}\right) - \Delta\mathcal{U}^m_{c}\left(\mathbf{m}\right)}{\mu_0M_s^2}
- \frac{1}{\varepsilon}\mathbf{n}\cdot\mathbf{m},
\end{equation}
where $\beta\mu_0M_s^2=2(-K_1+\Delta K_1)$, $K_1$, and $\Delta K_1$ are the cubic anisotropy constants at zero strain and the magnetostriction correction to it, respectively.
To obtain the equations determining $\mathbf m$, the conditions for the above minimum is used, which gives
\begin{equation}
m_x=\frac{n_x}{\lambda}-\varepsilon\left[\beta n_x^3-3\frac{\lambda_{100}n_x\sigma_{xx}+\lambda_{111}(n_y\sigma_{yy}+n_z\sigma_{zz})}{\mu_0M_s^2}\right]+O(\varepsilon^2),
\label{solut}
\end{equation}
here $\lambda=1+O(\varepsilon)$ is the Lagrange multiplier and similar equations hold for the components $m_{y,z}$.
The measurable quantity is the magnetization projection $M_1=\mathbf M\cdot {\mathbf{n}}$ averaged over the all possible orientations of the individual grains. Averaging Eq.(\ref{solut}) yields
\begin{equation}
 \frac{\langle M_1\rangle}{M_s} = \left\langle\sqrt{1-|\mathbf{m}\times{\mathbf{n}}|^2}\right\rangle = 1-\frac{1}{2}\left\langle|\mathbf{m}\times{\mathbf{n}}|^2\right\rangle+O(\varepsilon^3).
\label{m_highH}
\end{equation}
To accomplish our task, we express the saturated in-crystallite (local) stress components via their non-zero components in the laboratory frame, $\sigma_{11}$ and $\sigma_{22}$, using the transformation presented by Eq.~(3), substitute the results into Eq.~(\ref{m_highH}) and average it over all the directions of $\mathbf{n}$ and the angle $\psi$. This procedure, up to leading terms in the magnetostriction parameters, results in the law of approach to the magnetic saturation
\begin{equation}
\langle M\rangle \simeq M_s\left[1-\frac{1}{105}\frac{H_A^2(\sigma)}{H^2}\right],\;H_{A}^2(\sigma)=2M_s^2\left[\beta^2
-\frac{3\beta}{\mu_0M_s^2}(2\sigma_{11}-\sigma_{22})\Delta \lambda\right], \;\Delta \lambda=\lambda_{100}-\lambda_{111}.
\label{MH_curve}
\end{equation}

\begin{figure}[h]
\begin{center}
\includegraphics[width=0.48\textwidth]{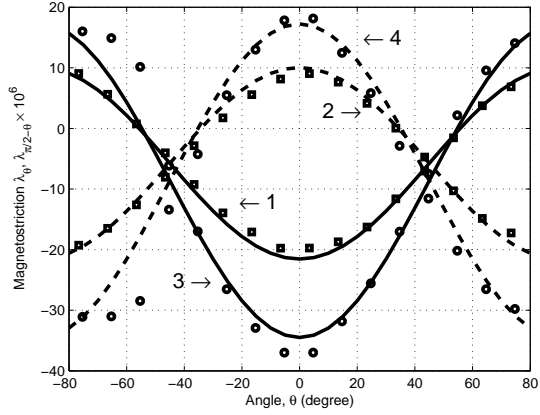}
\caption{\label{fig:sine} Magnetostriction as a function of angle $\theta$ for Ni and a Ni/PZT/Ni laminate. The curves are theoretical predictions for: solid -- $\lambda_{\theta}$, dashed - $\lambda_{\pi/2-\theta}$; $1$, $2$ -- the Ni/PZT/Ni laminate, and $3$, $4$ -- Ni. The markers represent the experimental data: squares -- Ni/PZT/Ni laminate, circles -- Ni.}
\end{center}
\end{figure}

\begin{figure}[h]
\begin{center}
\includegraphics[width=0.48\textwidth]{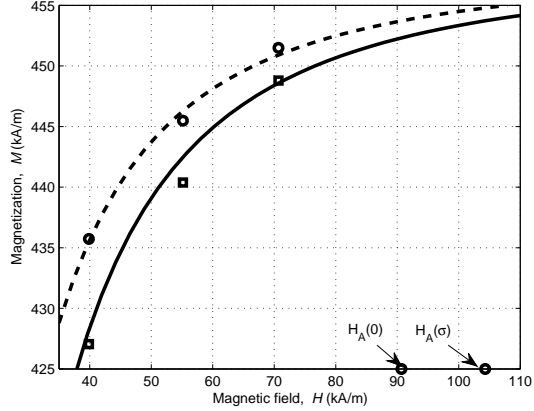}
\caption{\label{fig:MH} The magnetization curves in a high-$H$ range. The curves represent theoretical dependences and the markers represent the experimental data. The solid curve and squares correspond to the Ni/PZT laminate and the dashed curve and circles correspond to Ni  alone.}
\end{center}
\end{figure}

To verify the modeling of the inverse magnetostriction effect developed above, we bonded several laminates. Our raw materials were $99.95\%$ purity polycrystal Ni and APC-844 ceramic lead zinc titanate (PZT), obtained from Alfa Aesar and APC, respectively. Two $20\times 20$ mm$^2$ area samples with $0.5$ thick Ni and $0.83$ mm thick laminates were used for magnetostriction measurements, and another pair of $5\times 2$ mm$^2$ area samples with $0.25$ thick Ni and $0.58$ mm thick laminates, for magnetization measurements. The magnetostriction measurements were carried out using two-axis SR-4 strain gauges manufactured by Vishay Micro-Measurements. The magnetization curve at room temperature were measured on a PPMS machine.
Fig.~\ref{fig:sine} shows the saturated magnetostriction $\lambda_\theta$ and $\lambda_{\pi/2-\theta}$ vs $\theta$.  With material parameters  $E_{11}=76 ~\rm{GPa}$, $d_{31}=-109 ~\rm{pC/N}$, $\varkappa = 13.3~\rm{nF/m}$ (PZT), $E=160 ~\rm{GPa}$, $\nu=0.35$, and Eq.~(\ref{sigma_1&2&6}) we calculated the lamination stresses and used them for obtaining the theoretical curves. It is seen that a linear function of $\cos2\theta$ as predicted by Eq.(\ref{str}) fits good the measured dependences with the extracted  $\lambda_s=-3.4\times 10^{-5}$.
Fig.~\ref{fig:MH} shows a comparison of the calculated $\langle M\rangle$ vs $H$ curves in high-$H$ range against the experimental $M(H)$ data. For the laminate measured, we plugged into Eq.~(\ref{MH_curve}) the stresses $\sigma_{11} = 35~\rm{MPa} $ and $\sigma_{22}=-16~\rm{MPa}$ estimated using the numerical data \cite{zadov}. For the fitting parameters $\beta = 0.14 $, $M_s=458~\rm{kA/m}$ and $\Delta \lambda =-4.6\times 10^{-5}$ the theory agrees well with the experiment, as clearly seen from Fig.\ref{fig:MH}.

We have considered the inverse effect of magnetostriction in the ME laminates containing a magnetostrictive cubic ferromagnetic polycrystal. The lamination develops the stresses in  both the magnetostrictive and piezoelectric layer, and the stresses lead to the effective uniaxial magnetic anisotropy, see Eq.~(\ref{deltaU^m}). It is worth emphasizing that, similarly to ferromagnetic polycrystal \cite{landau,Chikazumi}, $\langle M\rangle$ of the laminates approaches the magnetic saturation in the inverse power-two law manner, while the lamination changes the value of the effective anisotropy field $H_A(\sigma)$, see Eq.~(\ref{MH_curve}). On the contrary, an initial susceptibility-adjusted Langevin curve \cite{Zheng,Li} for $\langle M\rangle$ vs $H$ predicts unrealistically fast exponential saturation.
For a Ni/PZT laminate, we have shown that, as compared to a bulk ferromagnetic, the magnetostriction of the laminate decreases, see Eq.~(\ref{str}) and Fig.~\ref{fig:sine}, and the magnetic anisotropy field increases,  see Eq.~(\ref{MH_curve}) and Fig.~\ref{fig:MH}.

The considered effect provides helpful insight into and adds additional feature to the mechanism of the strain-mediated ME coupling. Indeed in our notations, the thumb rule estimation of the ME coupling contains the factor $\left(\lambda_\theta - \lambda_{\pi/2-\theta}\right)/H_A$. For Ni as the magnetostrictive phase, the saturated magnetostriction and magnetic anisotropy field decreases and increases, respectively, in a concurrent manner. This explains why the measured ME coupling is smaller than the theoretical estimation of it obtained without including the inverse effect of magnetostriction, as noted in Introduction. Also in the present framework, the observed shift of the magnetic field of the optimal ME response to higher values relative to one estimated from the magnetization curve of the bulk ferromagnetic phase finds natural explanation. The outcomes of our theory and their experimental tests are reported here for a range near the magnetic saturation, but the opposite range of magnetic fields is treated similarly. Our study leads us to conclude that tailoring the stresses may essentially improve the strain-mediated ME coupling by compensating the inverse effect of magnetostriction. The account of this effect may also help for engineering the ME heterostructures with low magnetic bias and reduced adverse effects of domain structure, e.g. Barkhausen jumps.

The authors are grateful to I. Felner and M. Tsindlekht from Hebrew University for their help in the magnetization curves measurements.

\end{document}